\documentclass[
aps,
pra,
reprint,
superscriptaddress,
nofootinbib,
longbibliography
]{revtex4-2}

\usepackage[T1]{fontenc}
\usepackage{amsmath,amssymb,amsthm,mathtools,bm}
\usepackage{physics}
\usepackage{bbm}
\usepackage{graphicx}
\usepackage{xcolor}
\usepackage{hyperref}

\usepackage{float}
\makeatletter
\let\newfloat\newfloat@ltx
\makeatother

\usepackage{algorithm}
\usepackage{algpseudocode}

\usepackage{array}
\usepackage{booktabs}
\usepackage{multirow}
\usepackage{enumitem}
\usepackage{braket}
\usepackage{dsfont}
\usepackage{subcaption}
\usepackage{placeins}

\hypersetup{
    colorlinks=true,
    linkcolor=blue,
    citecolor=blue,
    urlcolor=blue
}

\theoremstyle{definition}

\theoremstyle{remark}

\begin{document}


\title{Hardware-Friendly Estimation of LCU-Based Expectation Values via Quantum-to-Classical Workload Shifting}

\author{Hari Hara Suthan Chittoor}
\email{hari.chittoor@qpiai.tech}
\affiliation{QpiAI India Private Limited, Bengaluru, India -- 560065}

\author{Thirumalai Manimaran}
\email{thirumalai.m@qpiai.tech}
\affiliation{QpiAI India Private Limited, Bengaluru, India -- 560065}


\begin{abstract}

Estimating the expectation value of an observable $O$ with respect to a quantum state evolved under a general, possibly non-unitary, operator $A$ is a fundamental task in quantum simulation, quantum chemistry, and variational quantum algorithms. Existing approaches based on the linear combination of unitaries (LCU) framework implement $A$ on quantum hardware, requiring ancilla qubits, controlled unitaries and, in randomized variants, a fresh set of circuit executions for every Monte Carlo sample. These requirements are costly on noisy intermediate-scale quantum (NISQ) and early fault-tolerant devices. We propose HERMES: \textbf{H}ardware-\textbf{E}fficient \textbf{R}eusable \textbf{M}easurements for \textbf{E}stimating LCU-based expectation value\textbf{S}, which offloads the quantum processing of $A$ entirely to the classical domain. By expanding both $A$ and $O$ in the Pauli basis and replacing the quantum evaluation of $A$ with classical Monte Carlo sampling over pairs of Pauli terms, HERMES rewrites $\mu = \mathrm{Tr}(OA\rho A^{\dagger})/\mathrm{Tr}(A\rho A^{\dagger})$ as a ratio of weighted sums of Pauli expectation values on the unmodified state $\rho$, requiring no ancilla qubits, controlled unitaries, or quantum implementation of $A$. A two-level measurement reduction scheme, combining Pauli deduplication and commuting-group partitioning, bounds the number of distinct measurement settings independently of the number of Monte Carlo samples, unlike randomized LCU methods such as single-ancilla LCU, whose circuit executions grow linearly with the sample count. We demonstrate HERMES on ground-state energy estimation of the $\mathrm{H}_2$ and $\mathrm{LiH}$ molecules via imaginary-time evolution, benchmarked against Pauli single-ancilla LCU baseline. HERMES achieves accuracy comparable to or better than the baseline with shallow circuits that, under qubit-wise commuting (QWC) grouping, contain no two-qubit gates beyond state preparation. It also shows greater robustness to hardware noise and completes the noisy $\mathrm{LiH}$ simulation that was infeasible for the baseline, which makes it attractive for NISQ devices, albeit with a classical overhead that grows with the number of Pauli terms in $A$.

\end{abstract}

\maketitle


\section{Introduction}

Quantum computing holds transformative promise for simulating quantum systems, solving linear systems, processing quantum information, and performing optimization tasks intractable for classical computers~\cite{feynman1982,shor1999polynomial,lloyd1996universal,Book_Osvaldo_Simeone_CQIT_2026}. A central computational primitive in many of these applications is estimating expectation values of the form
\[
\mu = \frac{\mathrm{Tr}(OA\rho A^\dagger)}{\mathrm{Tr}(A\rho A^\dagger)},
\]
where $\rho$ is a quantum state, $O$ is a Hermitian observable, and $A$ is a general possibly non-unitary operator. This setting covers Hamiltonian simulation ($A = e^{-iHt}$)~\cite{kassal2011simulating,berry2015simulating_truncated_Taylor_series}, ground-state estimation via imaginary-time evolution ($A = e^{-\tau H}$)~\cite{motta2020determining,lin2022heisenberg,dong2022ground}, and quantum linear systems ($A \propto B^{-1}$)~\cite{harrowHHL2009quantum}. In all cases, the desired output is the scalar $\mu$---not the full evolved state---a distinction central to this work.

The standard approach implements $A$ via the linear combination of unitaries (LCU) method~\cite{childsWiebe2012hamiltonian}, which requires $O(\log M)$ ancilla qubits and multi-qubit controlled gates. LCU is the foundational primitive for block encodings and quantum singular value transformation (QSVT)~\cite{gilyen_Wiebe2019QSVT_foundational_paper}, a framework of remarkable generality: by applying a polynomial transformation to the singular values of a block-encoded operator, QSVT subsumes quantum search, phase estimation, Hamiltonian simulation, and quantum linear system solvers~\cite{martyn2021grand_unification_of_quantum_algorithms_QSVT} --- thereby providing a unification of most known quantum algorithms, including Grover's search \cite{Grovers_algorithm_1996paper} and Shor's factoring \cite{Shors_algorithm_1994paper}. However, the ancilla register and multi-qubit controlled gate overhead of LCU place it out of reach on current noisy intermediate-scale quantum (NISQ)~\cite{preskill2018quantum,Book_Osvaldo_Simeone_QML} and early fault-tolerant~\cite{katabarwa2024early_FTQC, campbell2021early_FTQC_hubbard} hardware.

To reduce circuit complexity, \cite{chakraborty2024implementing_single_ancilla_LCU} introduced \emph{single-ancilla LCU}, which eliminates the multi-qubit ancilla register by randomizing over pairs of LCU terms and executing independent Hadamard-test circuits, each requiring only a single ancilla qubit. This shifts part of the workload from quantum to classical processing --- not to reduce total computational cost, but to make the quantum portion executable on current hardware where deep multi-qubit circuits are infeasible --- at the cost of increased total circuit executions. The trade-offs across the full coherent-to-randomized spectrum were studied in \cite{wada2026tradeoffsquantumclassicalresources}. Stochastic combination of unitaries was proposed in \cite{peetz2026quantumSimulationStochasticCombination} to reduce per-sample circuit depth. \cite{sun2025randomised_composite_LCU} extended randomized LCU to effective states of the form $U\rho V^\dagger$. Nevertheless, all randomized variants share a common requirement: the quantum hardware must coherently implement at least some unitaries in the LCU of $A$, and each Monte Carlo sample demands a fresh circuit execution. Consequently, for a fixed number of shots per sample, the total number of quantum circuit executions grows linearly with the sample count, and no measurements can be reused across samples.

A complementary and independently developed line of work addresses the measurement side of the problem. For an observable $O$ with $L$ Pauli terms, naive expectation estimation of $O$ requires up to $L$ distinct measurement settings. Pauli grouping methods reduce this overhead by partitioning the Pauli strings of $O$ into groups that can be measured simultaneously. Qubit-wise commuting (QWC) groups share a common tensor-product eigenbasis and therefore require only single-qubit rotations, with the minimization of the number of groups being cast as a minimum clique cover problem that is solved heuristically~\cite{verteletskyi2020measurementOptimizationinVQEminCliqueCover}. Partitioning into general mutually commuting groups reduces the number of measurement settings further, since each group can be simultaneously diagonalized by a Clifford circuit, at the cost of two-qubit gates in the rotation circuit~\cite{crawford2021efficientmeasurementofPaulioperatorsinfintiesamplingerror}; scalable partitioning algorithms exist for both QWC and general commuting groups~\cite{reggio2024fast_partition_Paulis}. A distinct strategy, unitary partitioning, instead reduces the number of terms to be estimated by appending additional coherent rotations to the state-preparation circuit~\cite{zhao2020measurementreductioninVQalgorithm}. However, this line of work has been developed primarily for estimating the expectation value of a fixed observable, and has not been integrated with LCU-based frameworks for estimating $\mu$ --- where the measurement structure is richer and depends jointly on the Pauli strings of both $A$ and $O$.

Despite the significant progress in both directions, a gap remains: to the best of our knowledge, no existing approach avoids ancilla qubits, controlled unitary gates, and any quantum implementation of $A$, thereby keeping the quantum circuits shallow and free of multi-qubit gates beyond state preparation. This gap is of practical importance, since on current NISQ hardware the fidelity of an estimate is limited primarily by circuit depth and by the number of two-qubit gates, whose error rates are substantially higher than those of single-qubit operations. Crucially, however, not every shift of workload from quantum to classical processing is beneficial: classical simulation of generic quantum states and circuits is believed to require resources that grow exponentially with the number of qubits~\cite{feynman1982,preskill2018quantum}. The key is to offload only those tasks that remain classically tractable, such as sampling over the Pauli expansion of $A$ when it contains a manageable number of terms, while retaining the quantum hardware for estimating the Pauli expectation values of $\rho$.

In this work, we present HERMES (\textbf{H}ardware-\textbf{E}fficient \textbf{R}eusable \textbf{M}easurements for \textbf{E}stimating LCU-based expectation value\textbf{S}), an algorithm that takes a fundamentally different approach by entirely offloading the quantum processing associated with $A$ to the classical domain. By expanding both $A$ and $O$ in the Pauli basis and replacing the direct quantum implementation of $A$ with classical Monte Carlo sampling over pairs of Pauli terms, drawn with probabilities proportional to the magnitudes of the corresponding coefficients of $A$, HERMES rewrites $\mu$ as a ratio of weighted sums of Pauli expectation values $\{\mathrm{Tr}(P\rho)\}$ evaluated on the unmodified input state $\rho$. The quantum hardware is never asked to implement $A$: its only tasks are state preparation, Clifford basis rotations for simultaneous measurement of groups of Pauli strings, and computational-basis measurement, with no ancilla qubits. This workload shift is classically tractable whenever $A$ has a manageable number of terms in the Pauli basis, as is the case for low-degree polynomial approximations of local Hamiltonians. Since $\mu$ is expressed entirely in terms of expectation values of Pauli strings that arise as products of strings from $A$ and $O$, the number of distinct Pauli expectation values, and hence of measurement settings, is bounded above independently of the Monte Carlo sample count. A two-level measurement reduction scheme --- Pauli deduplication followed by commuting-group partitioning --- further reduces the number of measurement settings, which therefore saturates as the sample count grows, whereas in single-ancilla LCU every sample requires a fresh circuit execution. With QWC grouping, two-qubit gates can be avoided entirely beyond state preparation, making HERMES particularly hardware-friendly.

The remainder of this paper is organized as follows. Section~\ref{sec:background} reviews the LCU framework, formulates the expectation value estimation problem, and describes the single-ancilla LCU method that serves as the baseline. Section~\ref{sec:proposed_algorithm} presents the proposed HERMES algorithm, including its classical preprocessing, quantum processing, and classical post-processing stages. Section~\ref{sec:experiments} reports numerical simulation results for ground-state energy estimation of the $\mathrm{H}_2$ and $\mathrm{LiH}$ molecules, benchmarked against the Pauli single-ancilla LCU (PSA-LCU) method. Section~\ref{sec:conclusion} concludes the paper and discusses the applicability of HERMES to other LCU-based problems together with open research directions.

\textbf{Notations:} $\mathbb{R}$ represents set of real numbers and $\mathbb{R}^{+}_{*}$ represents set of all real and positive numbers. $[M]$ denotes the set $\{1,2, \ldots, M\}$, where $M$ is a natural number.

\section{\label{sec:background}LCU Framework and Problem Formulation}

In this section, we review the LCU method and its hardware limitations, formulate the expectation value estimation problem, and describe the single-ancilla LCU algorithm that serves as the primary baseline for our work. We begin with some preliminaries on Pauli operators and commuting Pauli strings.

\subsection{Preliminaries}
\subsubsection{\label{subsec:pauli}Pauli Operators and Pauli Strings}

The single-qubit Pauli operators form a complete orthogonal basis for the space of $2\times2$  operators and are given by
\begin{equation*}
I=
\begin{bmatrix}
1&0\\
0&1
\end{bmatrix},
\,
X=
\begin{bmatrix}
0&1\\
1&0
\end{bmatrix},
\,
Y=
\begin{bmatrix}
0&-i\\
i&0
\end{bmatrix},
\,
Z=
\begin{bmatrix}
1&0\\
0&-1
\end{bmatrix}.
\end{equation*}
These operators satisfy
\begin{equation*}
X^2=Y^2=Z^2=I,
\end{equation*}
and obey the multiplication rules
\begin{equation*}
XY=iZ,\quad
YZ=iX,\quad
ZX=iY,
\end{equation*}
with reversed order introducing a negative phase,
\begin{equation*}
YX=-iZ,\quad
ZY=-iX,\quad
XZ=-iY.
\end{equation*}

An $n$-qubit Pauli string is the tensor product of single-qubit Pauli operators,
\begin{equation*}
P=P^{(1)}\otimes P^{(2)}\otimes\cdots\otimes P^{(n)},
\end{equation*}
where $P^{(i)}\in\{I,X,Y,Z\},$ for every $i\in [n]$. The set of all Pauli strings form a basis for operators acting on an $n$-qubit Hilbert space, allowing any operator to be expressed as a linear combination of Pauli strings. The product of two Pauli strings is obtained by multiplying the corresponding single-qubit operators independently at each qubit, resulting in another Pauli string up to an overall phase factor of $\{\pm1,\pm i\}$ inherited from the single-qubit multiplication rules.

\subsubsection{\label{subsec:commuting}Commuting Pauli Operators}

Two Pauli strings $P_i$ and $P_j$ either commute or anticommute,
\begin{equation*}
P_iP_j=\pm P_jP_i.
\end{equation*}
They commute if the number of qubit positions at which the corresponding non-identity Pauli operators anticommute is even, and anticommute otherwise. Equivalently,
\begin{equation*}
[P_i,P_j]= P_i P_j - P_j P_i =0
\end{equation*}
for commuting pairs, while
\begin{equation*}
P_i P_j + P_j P_i =0
\end{equation*}
for anticommuting pairs.

The commutation property is fundamental to measurement optimization in quantum algorithms \cite{reggio2024fast_partition_Paulis}. Since mutually commuting Pauli operators share a common eigenbasis, they can be measured simultaneously after an appropriate basis transformation. Consequently, commuting Pauli strings can be grouped into measurement sets, allowing multiple expectation values to be estimated from the outcomes of a single measurement setting and thereby reducing the total measurement cost.


\subsection{\label{subsec:lcu}Linear Combination of Unitaries}

An $n$-qubit operator $A$ can be expressed as a linear combination of $M$ unitaries,
\begin{equation}
    A \;=\; \sum_{j=1}^{M} a_j e^{i \theta_j} U_j,
    \label{eq:lcu_general}
\end{equation}
where $a_j \in \mathbb{R}_{*}^{+}$, $-\pi < \theta_j \leq \pi$, and $\{U_j\}$ are unitary operators. Without loss of generality, the phase terms can be absorbed into the unitaries as $\widetilde{U_j} = e^{i\theta_j} U_j$. Defining the one-norm $\lambda = \sum_{j=1}^{M} a_j$, the normalized coefficients $p_j = a_j/\lambda$ form a probability distribution over $[M]$, giving
\begin{equation}
\label{eq:lcu_normalized}
    A \;=\; \lambda \sum_{j=1}^{M} p_j \widetilde{U_j}.
\end{equation}
Such decompositions exist for all operators of physical interest and can be computed efficiently via dedicated classical algorithms~\cite{koska2024tree_approach_for_Pauli_decomposition,georges2025pauli_decomposition_via_Fast_WalshHadamard_transform,hantzko2024tensorizedPaulidecomposition}.

\begin{figure}[htbp]

\begin{subfigure}[htbp]{0.48\textwidth}
    \includegraphics[width=\linewidth]{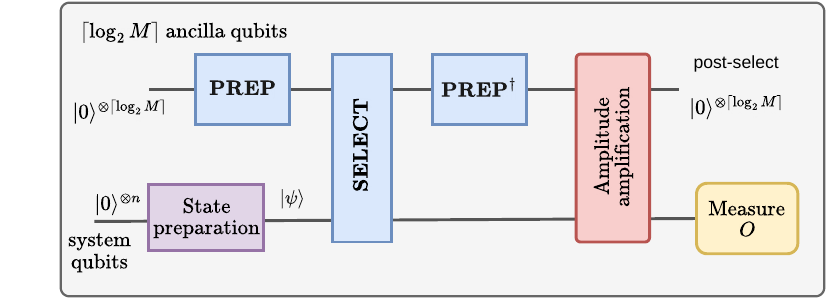}
    \caption{Standard LCU}
    \label{fig:standard_LCU}
\end{subfigure}
\hfill
\begin{subfigure}[htbp]{0.48\textwidth}
    \includegraphics[width=\linewidth]{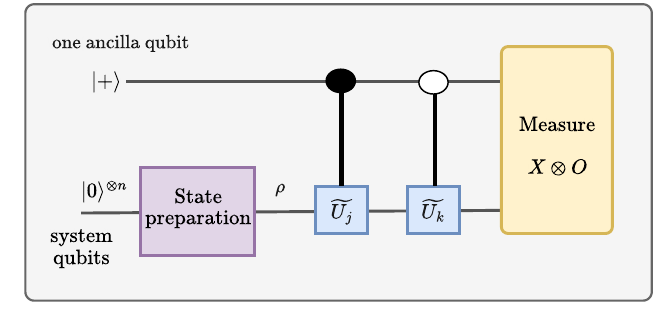}
    \caption{Single-ancilla LCU}
    \label{fig:single_ancilla_LCU}
\end{subfigure}
\hfill
\begin{subfigure}[htbp]{0.48\textwidth}
    \includegraphics[width=\linewidth]{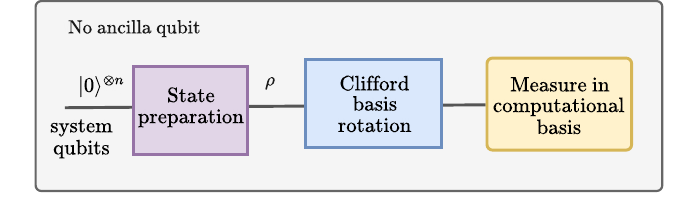}
    \caption{HERMES}
    \label{fig:HERMES}
\end{subfigure}

\caption{
Quantum circuits of LCU-based expectation value estimation methods. (a) Standard LCU \cite{childsWiebe2012hamiltonian} requires $\lceil \log_2 M \rceil$ ancilla qubits, PREP/SELECT oracles, and amplitude amplification across all qubits. (b) Single-ancilla LCU \cite{chakraborty2024implementing_single_ancilla_LCU} reduces to a single ancilla and two controlled unitaries per Monte Carlo sample, at the cost of a fresh circuit execution per sample. (c) HERMES requires only state preparation and Clifford rotations (single-qubit Clifford gates for QWC grouping), with no ancilla qubits.
}

\label{fig:comparison_of_quantum_circuits}

\end{figure}

We now describe the standard LCU procedure using pure states $|\psi\rangle$; extension to mixed states is straightforward. To realize $|\psi\rangle \mapsto A|\psi\rangle$ coherently, the procedure employs two oracles on the $n$-qubit target system $S$ and an ancilla register $\mathcal{A}$ of $\lceil \log_2 M \rceil$ qubits~\cite{childsWiebe2012hamiltonian,berry2015simulating_truncated_Taylor_series,low2019hamiltonian}. The \textsc{prep} oracle encodes the coefficient distribution,
\begin{equation}
    \textsc{prep}\;|0\rangle_{\mathcal{A}}
    \;=\; \sum_{j=1}^{M}\sqrt{p_j}\,|j\rangle_{\mathcal{A}},
    \label{eq:prepare}
\end{equation}
and the \textsc{select} oracle applies the $j$-th unitary conditioned on the ancilla,
\begin{equation}
    \textsc{select}
    \;=\; \sum_{j=1}^{M}|j\rangle\langle j|_{\mathcal{A}} \otimes \widetilde{U_j}.
    \label{eq:select}
\end{equation}
The combined circuit $U_{LCU} = (\textsc{prep}^\dagger \otimes I_S) \cdot \textsc{select} \cdot (\textsc{prep} \otimes I_S)$ satisfies the LCU lemma~\cite{childsWiebe2012hamiltonian,an2023LCUlemma_Hamiltoniansimulation}: for any state $|\psi\rangle_S$,
\begin{equation}
    U_{LCU}\,|0\rangle_{\mathcal{A}}|\psi\rangle_S
    \;=\; |0\rangle_{\mathcal{A}}
    \!\left(\sum_{j=1}^{M}p_j \widetilde{U_j}\right)\!|\psi\rangle_S
    + |\perp_\psi\rangle,
    \label{eq:lcu_lemma}
\end{equation}
where $|\perp_\psi\rangle$ is orthogonal to $|0\rangle_{\mathcal{A}} \otimes \mathcal{H}_S$ and $\mathcal{H}_S$ is the Hilbert space of the system qubits. Amplitude amplification boosts the probability of obtaining the ancilla outcome $|0\rangle_{\mathcal{A}}$, after which post-selecting $|0\rangle_{\mathcal{A}}$ on the ancilla register prepares $A|\psi\rangle$ (up to normalization) on the system register, and any observable $O$ can be measured directly on the prepared state. The standard LCU is depicted in Fig. \ref{fig:standard_LCU}. The hardware constraints of this procedure---$O(\log M)$ ancilla qubits, multi-qubit controlled-$\widetilde{U_j}$ gates, and amplitude amplification rounds multiplying the effective circuit depth---motivate the hardware-efficient alternatives discussed in Section~\ref{subsec:single_ancilla}.


\subsection{\label{subsec:problem}Expectation Value Estimation under Operator Evolution}

Consider an $n$-qubit input state $\rho$, a general (possibly non-unitary) operator $A$, and a Hermitian observable $O$. The normalized state evolved under $A$ is
\begin{equation}
    \tilde{\rho}
    \;=\;
    \frac{A\rho A^\dagger}{\mathrm{Tr}(A\rho A^\dagger)},
    \label{eq:evolved_state}
\end{equation}
and the target quantity is
\begin{equation}
\label{eq:mu}
    \mu
    \;=\;
    \mathrm{Tr}(O\tilde{\rho})
    \;=\;
    \frac{\mathrm{Tr}(OA\rho A^\dagger)}{\mathrm{Tr}(A\rho A^\dagger)}
    \;=:\;
    \frac{\mu_N}{\mu_D},
\end{equation}
where
\begin{equation}
\label{eq: mu_N}
    \mu_N = \mathrm{Tr}(OA\rho A^\dagger)
\end{equation}
and
\begin{equation}
\label{eq: mu_D}
    \mu_D = \mathrm{Tr}(A\rho A^\dagger)
\end{equation}
are the numerator and denominator, respectively. When $A$ is unitary, $\mu_D = 1$ and the problem reduces to standard expectation value estimation. For non-unitary $A$, the normalization factor $\mu_D$ can be exponentially small in the system size, making ratio estimation statistically challenging. A critical observation is that coherently preparing $\tilde{\rho}$ on quantum hardware is unnecessary for estimating $\mu$: the standard LCU pipeline pays the full cost of the ancilla register, \textsc{prep}--\textsc{select} circuit, and amplitude amplification even though the desired output is merely a scalar. This motivates methods that estimate $\mu_N$ and $\mu_D$ directly without ever preparing $\tilde{\rho}$, treating the quantum computer only as a device for measuring expectation values of observables. The single-ancilla LCU method reviewed next is a step in this direction; and the proposed HERMES algorithm in Section \ref{sec:proposed_algorithm} goes further by eliminating quantum evaluation of $A$ entirely.

\subsection{\label{subsec:single_ancilla}Single-Ancilla LCU}

The \emph{single-ancilla LCU} method, proposed in \cite{chakraborty2024implementing_single_ancilla_LCU}, estimates $\mu_N$ and $\mu_D$ without preparing $\tilde{\rho}$, using only a single ancilla qubit rather than the multi-qubit ancilla register required by the standard LCU construction. The starting point is the expansion
\begin{equation}
    \mu_N \;=\; \lambda^2\!\sum_{j,k=1}^{M} p_j p_k\,\mathrm{Re}\!\left(\mathrm{Tr}\!\left[O\,\widetilde{U_j}\rho \widetilde{U_k}^\dagger\right]\right),
    \label{eq:single_ancilla_expand}
\end{equation}
which follows from substituting~\eqref{eq:lcu_normalized} in \eqref{eq: mu_N} and using linearity of the trace. Each term $\mathrm{Re}(\mathrm{Tr}[O\,\widetilde{U_j}\rho  \widetilde{U_k}^\dagger])$ is estimated by a Hadamard-test-type circuit: a single ancilla qubit is initialized to $|+\rangle = (|0\rangle+|1\rangle)/\sqrt{2}$, $\widetilde{U_j}$ and $\widetilde{U_k}$ are applied to $\rho$ conditioned on the ancilla being in $|1\rangle$ and $|0\rangle$ respectively, followed by measurement of $X \otimes O$. The denominator $\mu_D$ is estimated analogously with $O = I$ in~\eqref{eq:single_ancilla_expand}, and $\mu = \mu_N/\mu_D$ is obtained by dividing the two estimates.

Although the per-sample circuit is substantially shallower than the standard LCU~\cite{chakraborty2024implementing_single_ancilla_LCU}, the key hardware demands remain. The quantum hardware must coherently apply two controlled unitaries $\widetilde{U_j}$ and $\widetilde{U_k}$ per sample and measure the composite observable $X \otimes O$. In practice, since $O$ is not directly measurable as a single quantum observable, it is first expanded as a linear combination of Pauli terms; the Hadamard-test-type circuit is then re-executed once per commuting group of $O$, so each Monte Carlo sample already requires at least $G_O$ circuit executions, where $G_O$ is the number of commuting groups of $O$. Furthermore, a fresh set of such circuits must be executed for every new pair $(j,k)$, with no mechanism to reuse measurements across samples. Consequently, the total quantum execution cost grows as $\mathcal{O}(R \cdot G_O)$ with the Monte Carlo sample count $R$, with no saturation. Single-ancilla LCU is depicted in Fig.~\ref{fig:single_ancilla_LCU}. In the next section we present HERMES, which overcomes all of these limitations.


\section{Proposed Algorithm}
\label{sec:proposed_algorithm}

In this section, we present the HERMES algorithm to compute $\mu$ in (\ref{eq:mu}). The complete framework consists of three major stages: classical preprocessing, quantum processing, classical post-processing.

\subsection{Classical Preprocessing}

The objective of the classical preprocessing stage is to transform the original non-unitary expectation estimation problem into estimation of a structured collection of Pauli observables. This stage is entirely classical and does not require quantum hardware execution. The classical preprocessing stage consists of the following components:

\subsubsection{Pauli Decomposition of operator $A$ and observable $O$}

We begin by expanding the generally non-unitary operator $A$ as a LCU, with the complete $n$-qubit Pauli basis as the unitary basis:
\begin{equation}
    A \;=\; \sum_{j=1}^{M} a_j e^{i\theta_j} P_j,
    \label{eq:A_decomp}
\end{equation}
where $P_j \in \{I,X,Y,Z\}^{\otimes n}$ are Pauli strings, $a_j \in \mathbb{R}^{+}_{*}$, and $\theta_j \in (-\pi, \pi]$. As $A$ is an arbitrary operator, the coefficients $a_j e^{i\theta_j}$ are in general complex numbers. Since the Pauli operators form a complete orthogonal operator basis under the Hilbert--Schmidt inner product,
\begin{equation*}
\operatorname{Tr}(P_iP_j) = 2^n \delta_{ij},
\end{equation*}
where $\delta_{ij}$ is the Kronecker delta function, any $n$-qubit operator admits an exact expansion in this basis.

The Hermitian observable $O$ is similarly expanded as
\begin{equation}
O = \sum_{l=1}^{L} b_l P_l,
\label{eq:O_decomposition}
\end{equation}
where $b_l\in\mathbb{R}\setminus \{0\}$. Let $\mathcal{P}_A$ and $\mathcal{P}_O$ be the sets of all $M$ and $L$ Pauli strings in the LCU decompositions in (\ref{eq:A_decomp}) and (\ref{eq:O_decomposition}) respectively. Now using (\ref{eq:A_decomp}) and (\ref{eq:O_decomposition}), the numerator in (\ref{eq: mu_N}) can be written as
\begin{align}
\mu_N &= \sum_{j,k} a_j a_k e^{i(\theta_j -\theta_k)} \operatorname{Tr}(O P_j \rho P_k) \nonumber \\
&= \sum_{j,k} a_j a_k e^{i(\theta_j - \theta_k)} \sum_{l} b_l  \Tr(P_k P_l P_j \rho) \nonumber \\
&= \sum_{j,k} a_j a_k e^{i(\theta_j - \theta_k)} \sum_{l} b_l c_{klj} \Tr(P_{klj} \rho) \label{eq: mu_N expansion}
\end{align}
where $P_k P_l P_j = c_{klj} P_{klj}$ with $c_{klj} \in \{\pm 1, \pm i\}$. Similarly, using  (\ref{eq:A_decomp}) the denominator in (\ref{eq: mu_D}) can be written as
\begin{align}
\mu_D &= \sum_{j,k} a_j a_k e^{i(\theta_j -\theta_k)} \operatorname{Tr}( P_j \rho P_k) \nonumber \\
&= \sum_{j,k} a_j a_k e^{i(\theta_j - \theta_k)} \Tr(P_k P_j \rho) \nonumber \\
&= \sum_{j,k} a_j a_k e^{i(\theta_j - \theta_k)}c_{kj} \Tr(P_{kj} \rho) \label{eq: mu_D expansion}
\end{align}
where $P_k P_j = c_{kj} P_{kj}$ with $c_{kj} \in \{\pm 1, \pm i\}$.

These expressions reveal that the entire estimation problem reduces to finding expectation values of Pauli observables with respect to the state $\rho$. However, direct deterministic evaluation becomes computationally expensive since the numerator (\ref{eq: mu_N expansion}) contains $\mathcal{O}(M^2L)$ Pauli combinations and the denominator (\ref{eq: mu_D expansion}) contains $\mathcal{O}(M^2)$ Pauli combinations. To overcome this, we employ a Monte Carlo sampling strategy.

\subsubsection{Construction of probability distribution and Monte Carlo sampling}

Form a probability distribution using the coefficients of $A$ in (\ref{eq:A_decomp}) as

\begin{equation}
\Bigl\{p_j = \frac{a_j}{\lambda} : j \in [M]\Bigr\}
\label{eq:sampling_distribution}
\end{equation}
where $\lambda = \sum_{j=1}^{M} a_j$ is the $l_1$ norm of the coefficients of $A$.

Construct a set $\mathcal{R}$ of $R$ pairs of Monte Carlo samples, where each pair contains two indices sampled independently from the probability distribution 
$\{p_j: j\in[M]\}$. Thus $\mathcal{R} = \{(j,k):  j,k\sim\{p_j\}\}$.

\subsubsection{Generation of numerator and denominator Pauli observables}

For each unique sample $(j,k) \in \mathcal{R}$ and $l \in [L]$, we collect the Pauli strings $P_{klj}, P_{kj}$ and coefficients $c_{klj}, c_{kj}$ into the following four indexed collections, where each entry is uniquely identified by its index tuple $(k,l,j)$ or $(k,j)$ but the same Pauli string or coefficient value may appear under multiple distinct indices:

{\footnotesize
\begin{align*}
    \mathcal{P}_N \hspace{-0.05cm} &= \hspace{-0.05cm}\{P_{klj}: P_{k}P_{l}P_{j} = c_{klj} P_{klj}, (j,k) \in \mathcal{R}; P_l \in \mathcal{P}_O; P_j, P_k \in \mathcal{P}_A \} \\
    \mathcal{C}_N \hspace{-0.05cm} &= \hspace{-0.05cm} \{c_{klj}: P_{k}P_{l}P_{j} = c_{klj} P_{klj}, (j,k) \in \mathcal{R}; P_l \in \mathcal{P}_O; P_j, P_k \in \mathcal{P}_A \}  \\
    \mathcal{P}_D &= \{P_{kj}: P_{k}P_{j} = c_{kj} P_{kj}, (j,k) \in \mathcal{R}, P_j, P_k \in \mathcal{P}_A\} \\
    \mathcal{C}_D &= \{c_{kj}: P_{k}P_{j} = c_{kj} P_{kj}, (j,k) \in \mathcal{R}, P_j, P_k \in \mathcal{P}_A\}.
\end{align*}
}

Define $Z^{jk}_N$ and $Z^{jk}_D$ as 
\begin{align}
    Z^{jk}_N &=  e^{i(\theta_j - \theta_k)} \sum_{l} b_l c_{klj} \Tr(P_{klj} \rho) \label{eq: Z^jk_N} \\
    Z^{jk}_D &=  e^{i(\theta_j - \theta_k)} c_{kj} \Tr(P_{kj} \rho) \label{eq: Z^jk_D}
\end{align}

As $\mu_N$ and $\mu_D$ are real quantities we define the corresponding estimators as $\mathrm{Re}(Z^{jk}_N)$ and $\mathrm{Re}(Z^{jk}_D)$. Now consider the expectation value of $\mathrm{Re}(Z^{jk}_N)$ and $\mathrm{Re}(Z^{jk}_D)$
\begin{align*}
    \mathbb{E}_{j,k} [\mathrm{Re}(Z^{jk}_N)] &= \sum_{j,k} p_j p_k \mathrm{Re}(Z^{jk}_N) \\
    &= \scalebox{1}{$\sum\limits_{j,k} \frac{a_j}{\lambda} \frac{a_k}{\lambda} \mathrm{Re}\Bigl(\sum\limits_{l} b_l e^{i(\theta_j - \theta_k)}c_{klj} \Tr(P_{klj} \rho)\Bigr)$} \\
    &= \frac{\mu_N}{\lambda^2} \\
    \mathbb{E}_{j,k} [\mathrm{Re}(Z^{jk}_D)] &= \sum_{j,k} p_j p_k \mathrm{Re}(Z^{jk}_D) \\
    &= \sum_{j,k} \frac{a_j}{\lambda} \frac{a_k}{\lambda} \mathrm{Re}\bigl( e^{i(\theta_j - \theta_k)} c_{kj} \Tr(P_{kj} \rho)\bigr) \\
    &= \frac{\mu_D}{\lambda^2}
\end{align*}
Therefore, $\lambda^2 \mathrm{Re}(Z^{jk}_N)$ and $\lambda^2 \mathrm{Re}(Z^{jk}_D)$ are unbiased estimators of $\mu_N$ and $\mu_D$ respectively. Thus, the problem of estimating $\mu_N$ and $\mu_D$ is reduced to the problem of finding the expectation of Pauli observables in $\mathcal{P}_{N} \cup \mathcal{P}_{D}$. We need the quantum hardware only to compute expectation values of these Pauli observables. We emphasize that the operator $A$ or the corresponding Pauli strings in its LCU expansion are not directly implemented on the quantum hardware.


\begin{algorithm}[t]
\caption{Classical Preprocessing}
\label{alg:classicalpreprocessing}
\begin{algorithmic}[1]
\State \textbf{Input:} Square operator $A$, Hermitian observable $O$ on $n$ qubits and Number of samples $R$

\State \textbf{Compute LCU decomposition of operator $A$ and observable $O$ in the Pauli basis:} 

\noindent $A:  \{(a_j e^{i \theta_j}, P_j): j\in [M]\}$  where $a_j \in \mathbb{R}^{+}_{*}$, $\theta_j \in (-\pi, \pi]$

$O: \{(b_l , P_l): l\in [L]\}$ where $b_l \in \mathbb{R} \setminus \{0\}$


\State Initialize $\mathcal{P}_A$ and $\mathcal{P}_O$ with set of all Pauli strings in the decomposition of operator $A$ and observable $O$ respectively

\State \textbf{Probability distribution:} Form a probability distribution using the coefficients of decomposition of $A$ as $\{p_j = \frac{a_j}{\lambda} : j \in [M], \lambda=\sum\limits_{j=1}^{M}a_j\}$

\State \textbf{Monte Carlo sampling:} Construct a set $\mathcal{R}$ of $R$ pairs of samples, where each pair contains two indices sampled independently from the probability distribution $\{p_j: j\in[M]\}$. Thus $\mathcal{R} = \{(j,k):  j,k\sim\{p_j\}\}$

\State \textbf{Constructing $\mathcal{P}_N, \mathcal{P}_D, \mathcal{C}_N, \mathcal{C}_D$:}
\State Initialize $4$ empty indexed collections $\mathcal{P}_N, \mathcal{P}_D, \mathcal{C}_N$ and $\mathcal{C}_D$. $\mathcal{P}_N, \mathcal{P}_D$ will store Pauli strings and $ \mathcal{C}_N, \mathcal{C}_D$ will store coefficients $\{\pm 1, \pm i\}$.

\For{each unique $(j,k) \in \mathcal{R}$}

\For{each $l \in [L]$}
\State Compute $P_{k}P_{l}P_{j} = c_{klj} P_{klj}$, where $P_j, P_k \in \mathcal{P}_A; P_l \in \mathcal{P}_O$ and $c_{klj} \in \{\pm 1, \pm i\}$
\State Append $P_{klj}$ to $\mathcal{P}_N$ and $c_{klj}$ to $\mathcal{C}_N$.
\EndFor

\State Compute $P_{k}P_{j} = c_{kj} P_{kj}$, where $P_j, P_k \in \mathcal{P}_A$ and $c_{kj} \in \{\pm 1, \pm i\}$
\State Append $P_{kj}$ to $\mathcal{P}_D$ and $c_{kj}$ to $\mathcal{C}_D$.

\EndFor

\State \textbf{First Level of Measurement Reduction via Pauli deduplication:} Define $\mathcal{P}_{unique}$ as the set of distinct Pauli strings in $ \mathcal{P}_N \cup \mathcal{P}_D$, so that the expectation values only need to be computed for the elements of $\mathcal{P}_{unique}$ 

\State \textbf{Second Level of Measurement Reduction via groups of commuting Paulis:}
Partition $\mathcal{P}_{unique}$ into $G$ groups $\{\mathcal{P}_g: g \in [G]\}$ using a Pauli grouping strategy, where all Paulis in each group commutes with each other, such that a single measurement setting suffices to estimate the expectation values of all Pauli observables within each group.

\State \textbf{Output:} LCU Decomposition of $A$: $\{(a_j e^{i \theta_j}, P_j): j\in [M]\}$, and $O$: $\{(b_l , P_l): l\in [L]\}$. $G$ groups of commuting Paulis $\{\mathcal{P}_g: g \in [G]\}$. Sets $\mathcal{R}, \mathcal{P}_{unique}$ and collections $\mathcal{P}_N, \mathcal{C}_N, \mathcal{P}_D, \mathcal{C}_D$

\end{algorithmic}
\vspace{-0.1cm}

\end{algorithm}


\subsubsection{Two level measurement reduction}

As $R,M$ and $L$ increase, the number of Pauli observable expectations to be computed increases. To reduce this, we propose a two level measurement reduction scheme.

\paragraph{First level -- Pauli deduplication.}
Two different samples $(j,k)$ and $(j^{'},k^{'})$ in $\mathcal{R}$ may correspond to same Pauli string, i.e., $P_{klj}=P_{k^{'}l^{'}j^{'}}$ or $P_{klj}=P_{k^{'}j^{'}}$ or $P_{kj}=P_{k^{'}j^{'}}$, where $l,l^{'} \in [L]$. Thus, $\mathcal{P}_N \cup \mathcal{P}_D$ may contain many duplicates. By deduplication, we remove these repetitions, so that the expectation value of each distinct Pauli observable needs to be computed only once on quantum hardware, regardless of how many samples reference it. This is the first level of measurement reduction. Towards this end, we define 
\begin{equation}
    \mathcal{P}_{\mathrm{unique}} \;=\; \mathrm{Unique}\!\left(\mathcal{P}_N \cup \mathcal{P}_D\right).
    \label{eq:dedup}
\end{equation}

\paragraph{Second level -- commuting-group partitioning.}

We partition $\mathcal{P}_{\mathrm{unique}}$ into $G$ groups of mutually commuting Pauli strings~\cite{reggio2024fast_partition_Paulis}:
\begin{equation}
    \mathcal{P}_{\mathrm{unique}} \;=\; \bigcup_{g=1}^{G} \mathcal{P}_g.
    \label{eq:grouping}
\end{equation}
All Pauli strings within a commuting group share a common eigenbasis and can be measured simultaneously with a single basis-rotation circuit followed by computational basis measurements. This substantially reduces total quantum measurement overhead. As $R$ grows, newly sampled products increasingly collide with already-seen Pauli strings, causing $G$ to saturate at a finite ceiling bounded above by the Pauli support of $A$ and $O$, and in practice further reduced by the sparsity of their decompositions. Beyond a characteristic sample count $R^*$, adding more Monte Carlo samples does not increase $G$. This stands in sharp contrast to the single-ancilla LCU approach, where each sampled pair $(j,k)$ requires a separate circuit evaluation, so that the quantum execution cost continues to grow with the number of Monte Carlo samples. The complete classical preprocessing stage is summarized in Algorithm~\ref{alg:classicalpreprocessing}.


\subsection{Quantum Processing}

Following classical preprocessing, the HERMES algorithm proceeds to the quantum processing stage. The purpose of this stage is to estimate the expectation values of all distinct Pauli observables contained in $\mathcal{P}_{\mathrm{unique}}$ using a finite number of quantum measurement configurations.

For each commuting group $\mathcal{P}_g$ where $g \in [G]$, the algorithm first computes a unitary transformation $V_g$ that simultaneously diagonalizes all Pauli operators within the group $\mathcal{P}_g$. Since all operators commute, such a common eigenbasis always exists. The quantum state $\rho$ is then prepared on quantum hardware, after which the basis-rotation unitary $V_g$ is applied. The rotated state is subsequently measured in the computational basis using a finite number of shots. From the resulting bitstring statistics, the expectation values of all Pauli observables belonging to that commuting group are simultaneously estimated through classical processing. Consequently, a single measurement setting yields information about multiple observables simultaneously, substantially reducing the quantum execution cost. Moreover, if qubit-wise commuting (QWC) grouping is used, at the second level of measurement reduction in classical preprocessing stage, the basis rotations $V_g$ consist solely of single-qubit Clifford gates, so no two-qubit entangling gates are needed beyond those used for state preparation. The complete quantum processing stage is summarized in Algorithm~\ref{algorithm quantum processing} and is depicted in Fig. \ref{fig:HERMES}.


\begin{algorithm}[htbp]
\caption{Quantum Processing}
\label{algorithm quantum processing}
\begin{algorithmic}[1]
\State \textbf{Input:} $G$ groups of commuting Paulis $\{\mathcal{P}_g: g \in [G]\}$ and input quantum state $\rho$

\For{each $g \in [G]$}
\State Classically compute a unitary $V_g$ that simultaneously diagonalizes all Pauli operators in the commuting set $\mathcal{P}_g$

\State Prepare the input quantum state $\rho$ using state preparation circuit on quantum hardware
\State Apply the basis-rotation circuit corresponding to $V_g$ on the quantum state $\rho$

\State Perform computational basis measurements on the rotated state $V_g \rho V_g^\dagger$ using a finite number of shots

\State From the measurement outcomes, estimate the expectation values $\{\Tr(P\rho): P \in \mathcal{P}_g\}$ simultaneously using classical processing

\EndFor

\State \textbf{Output:} The expectation values $\{\Tr(P \rho): P \in \mathcal{P}_{unique} \}$

\end{algorithmic}
\end{algorithm}


\subsection{Classical Post-processing}

In the final stage of the HERMES algorithm, the results from the classical preprocessing and quantum processing stages are post-processed to estimate the desired non-unitary expectation value.

For each sample pair $(j,k)\in\mathcal{R}$, the algorithm computes $Z_N^{jk}$ and $Z_D^{jk}$ using (\ref{eq: Z^jk_N}) and (\ref{eq: Z^jk_D}). Since physical expectation values are real-valued, only the real components contribute to the final estimator. The numerator and denominator estimators are therefore constructed as
\begin{equation}
\hat{\mu}_N = \frac{1}{R}\sum_{(j,k)\in\mathcal{R}}\mathrm{Re}(Z_N^{jk}),
\end{equation}
and
\begin{equation}
\hat{\mu}_D = \frac{1}{R}\sum_{(j,k)\in\mathcal{R}}\mathrm{Re}(Z_D^{jk}).
\end{equation}

The final estimator is then given by
\begin{equation}
\hat{\mu} = \frac{\hat{\mu}_N}{\hat{\mu}_D}.
\label{eq:final_estimator}
\end{equation}

To further improve computational efficiency, the algorithm reuses previously computed quantities whenever duplicate Monte Carlo index pairs appear in the sampled set $\mathcal{R}$. This avoids recomputation and reduces the overhead of classical reconstruction. The complete postprocessing stage is summarized in Algorithm~\ref{algorithm classical postprocessing}.


\begin{algorithm}[htbp]
\caption{Classical Post-processing}
\label{algorithm classical postprocessing}
\begin{algorithmic}[1]
\State \textbf{Input:} Expectation of Pauli observables $\{\Tr(P \rho): P \in \mathcal{P}_{unique}\}$. Sets $\mathcal{R}, \mathcal{P}_{unique}$ and collections $\mathcal{P}_N, \mathcal{C}_N, \mathcal{P}_D, \mathcal{C}_D$. Decomposition of  $A: \{(a_j e^{i \theta_j}, P_j): j\in [M]\}$ and Decomposition of  $O: \{(b_l , P_l): l\in [L]\}$

\State Initialize $S_N = 0$ and $S_D = 0$ 
\For{each $(j,k) \in \mathcal{R}$}

\If{$Z^{jk}_N$ and $Z^{jk}_D$ are already computed}
\State $S_N \leftarrow S_N + \mathrm{Re}(Z^{jk}_N) $ and $S_D \leftarrow S_D + \mathrm{Re}(Z^{jk}_D) $

\Else
\State Initialize $Z^{jk}_N = 0$ and $Z^{jk}_D = 0$

\For{each $l \in [L]$}

\State $Z^{jk}_N \leftarrow Z^{jk}_N + b_l e^{i(\theta_j - \theta_k)}c_{klj} \Tr(P_{klj} \rho)$

\EndFor

\State $Z^{jk}_D = e^{i(\theta_j - \theta_k)}c_{kj} \Tr(P_{kj} \rho)$

\State $S_N \leftarrow S_N + \mathrm{Re}(Z^{jk}_N) $ and $S_D \leftarrow S_D + \mathrm{Re}(Z^{jk}_D) $
\EndIf
\EndFor

\State $\hat{\mu}_N \leftarrow S_N/R$ 
\State $\hat{\mu}_D \leftarrow S_D/R$

\State \textbf{Output:} $\hat{\mu} = \hat{\mu}_N / \hat{\mu}_D$

\end{algorithmic}
\end{algorithm}


Overall, HERMES transforms the estimation of non-unitary expectation values into a classical-quantum cooperative estimation problem, where the quantum hardware is responsible only for measuring Pauli observables on the unmodified state $\rho$ and all structure associated with $A$ is handled classically. By combining Monte Carlo sampling, Pauli deduplication, commuting-group partitioning, and cached reconstruction of repeated sample contributions, HERMES substantially reduces both the number of distinct quantum measurement settings and the classical postprocessing overhead. The method requires no ancilla qubits, no controlled unitary gates (with QWC grouping), no amplitude amplification, and no quantum evaluation of $A$ in any form, reducing the quantum circuit to state preparation and single-qubit Clifford basis rotations --- making HERMES directly executable on near-term devices where qubit count, circuit depth, and measurement cost are the dominant practical constraints.



\section{Results}
\label{sec:experiments}
 
\begin{figure*}[t]
    \centering
    \begin{subfigure}[b]{0.32\textwidth}
        \centering
        \includegraphics[width=\linewidth]{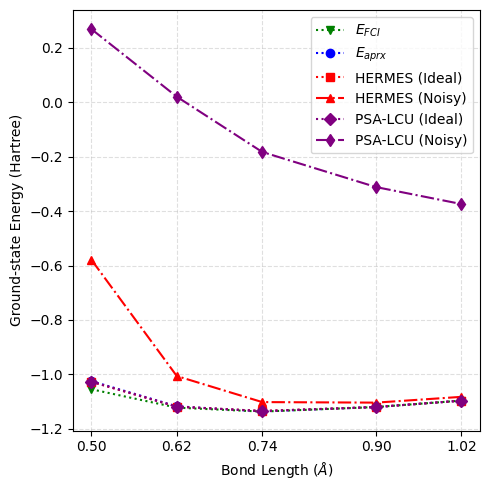}
        \caption{}
        \label{fig:h2_energy}
    \end{subfigure}
    \hfill
    \begin{subfigure}[b]{0.32\textwidth}
        \centering
        \includegraphics[width=\linewidth]{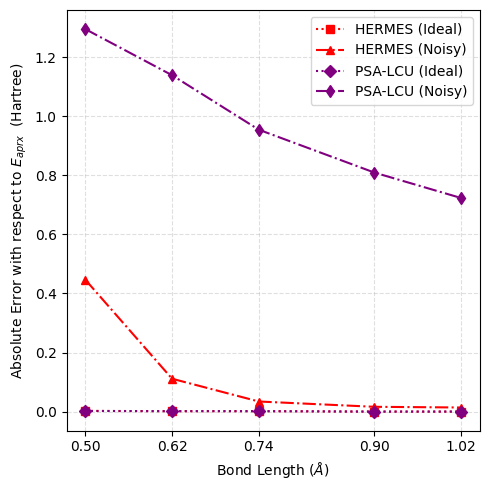}
        \caption{}
        \label{fig:h2_error}
    \end{subfigure}
    \hfill
    \begin{subfigure}[b]{0.32\textwidth}
        \centering
        \includegraphics[width=\linewidth]{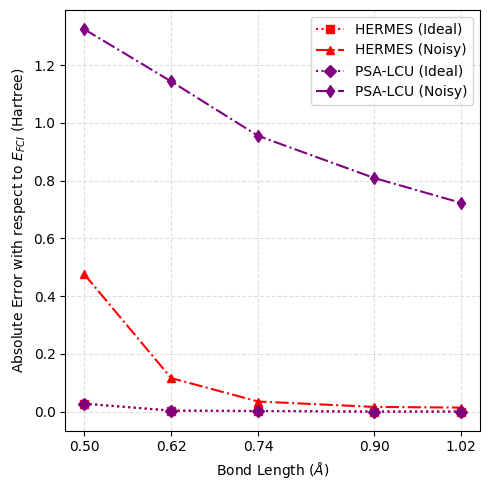}
        \caption{}
        \label{fig:h2_fci_error}
    \end{subfigure}
    \caption{Ground-state energy estimation for H$_2$: (a) $E_{aprx}$, $E_{FCI}$, and the ideal and noisy HERMES and PSA-LCU estimates (Hartree); (b) absolute error with respect to $E_{aprx}$; (c) absolute error with respect to $E_{FCI}$.}
    \label{fig:h2_results}
\end{figure*}
 
\begin{figure*}[t]
    \centering
    \begin{subfigure}[b]{0.32\textwidth}
        \centering
        \includegraphics[width=\linewidth]{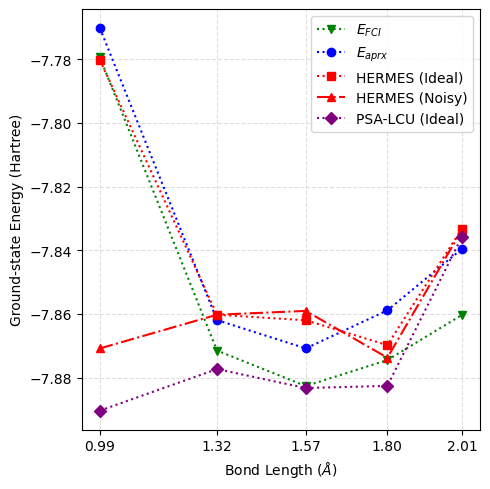}
        \caption{}
        \label{fig:lih_energy}
    \end{subfigure}
    \hfill
    \begin{subfigure}[b]{0.32\textwidth}
        \centering
        \includegraphics[width=\linewidth]{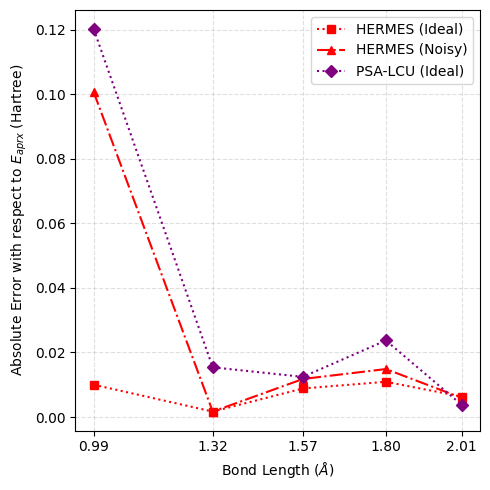}
        \caption{}
        \label{fig:lih_error}
    \end{subfigure}
    \hfill
    \begin{subfigure}[b]{0.32\textwidth}
        \centering
        \includegraphics[width=\linewidth]{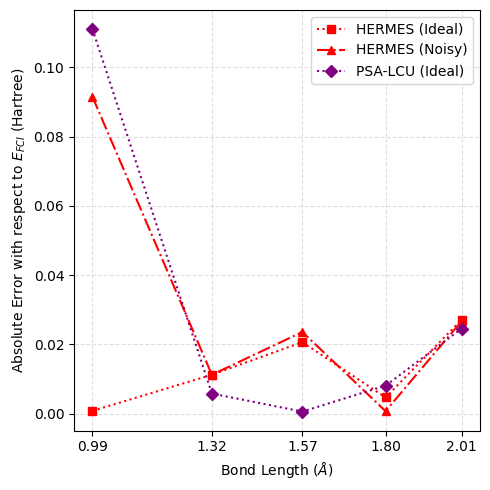}
        \caption{}
        \label{fig:lih_fci_error}
    \end{subfigure}
    \caption{Ground-state energy estimation for LiH: (a) $E_{aprx}$, $E_{FCI}$, the ideal and noisy HERMES estimates, and the ideal PSA-LCU estimate (Hartree); (b) absolute error with respect to $E_{aprx}$; (c) absolute error with respect to $E_{FCI}$. Noisy PSA-LCU results for $\mathrm{LiH}$ could not be obtained with the available resources (see Sec~\ref{subsec:lih} and Sec~\ref{subsec:discussion}).}
    \label{fig:lih_results}
\end{figure*}

\subsection{Benchmark and simulation setup}
\label{sec:setup}
 
The performance of \textsc{HERMES}, with QWC grouping, is assessed through numerical simulations of ground-state energy estimation for the $\mathrm{H}_2$ and $\mathrm{LiH}$ molecules, each evaluated at five bond lengths so that the behavior of the algorithm can be examined across the potential energy curve. Both molecules are described in the STO-3G basis. The qubit Hamiltonians contain $15$ Pauli terms on $4$ qubits for $\mathrm{H}_2$ and $631$ Pauli terms on $12$ qubits for $\mathrm{LiH}$, independent of the bond length. The molecular Hamiltonians and the full configuration interaction (FCI) energies $E_{FCI}$ are obtained from the PennyLane quantum chemistry dataset~\cite{Utkarsh2023ChemistryPennylaneDataset}. 
 
The results are benchmarked against the Pauli single-ancilla LCU (PSA-LCU) method, a special case of the single-ancilla LCU framework~\cite{chakraborty2024implementing_single_ancilla_LCU} in which every unitary in the LCU decomposition of $A$ is a Pauli string. Both methods therefore operate on an identical Pauli expansion of $A$, so that differences in performance can be attributed to the estimation strategy rather than to the decomposition. Implementing PSA-LCU requires one ancilla qubit in addition to the system register and controlled Pauli (two-qubit) gates, whereas \textsc{HERMES} uses neither: its circuits consist of state preparation and single-qubit basis rotations only.

\paragraph{Imaginary-time evolution.}
Ground-state estimation is carried out via imaginary-time evolution~\cite{motta2020determining,lin2022heisenberg,dong2022ground}. In contrast to real-time evolution, which is unitary, imaginary-time evolution is generated by the non-unitary propagator
\begin{equation}
    U(\tau) = e^{-\tau H},
    \label{eq:imaginary_time_propagator}
\end{equation}
where $H$ is the qubit Hamiltonian, expressed as a linear combination of Pauli strings, and $\tau$ is the imaginary-time parameter. Expanding an initial state in the eigenbasis of $H$ as $|\psi_0\rangle = \sum_k c_k |E_k\rangle$, with $E_0 < E_1 \le E_2 \le \cdots$, gives $e^{-\tau H}|\psi_0\rangle = \sum_k c_k e^{-\tau E_k}|E_k\rangle$. The amplitude of each excited eigenstate relative to that of the ground state is thus multiplied by $e^{-\tau (E_k - E_0)}$. For $c_0 \neq 0$, the normalized evolved state therefore approaches $|E_0\rangle$ as $\tau$ increases, and the energy expectation value
$\mathrm{Tr}(HA\rho A^\dagger)/\mathrm{Tr}(A\rho A^\dagger)$, with $A = U(\tau)$ and $\rho = |\psi_0\rangle\langle\psi_0|$, converges to $E_0$. This is precisely the form of expectation value $\mu$ in (\ref{eq:mu}) with $O = H$, which both \textsc{HERMES} and PSA-LCU estimate. In all simulations, the Hartree--Fock state is used as the initial state, since it has a substantial overlap with the ground state for the molecules considered.

\paragraph{Reference energies and error metrics.}
Since $e^{-\tau H}$ has no finite Pauli expansion, it is replaced by its second-order Taylor truncation,
\begin{equation}
    A_{aprx} = I - \tau H + \frac{\tau^2}{2}H^2,
    \label{eq:approx_A__imaginary_time}
\end{equation}
where $I$ is the identity operator and $H^2$ is expanded in the Pauli basis with like terms combined. The resulting operator $A_{aprx}$ contains $24$ Pauli terms for $\mathrm{H}_2$ and $25{,}542$ for $\mathrm{LiH}$. For all the simulations we set $\tau = 100$. 
The energy obtained from this operator,
\begin{equation}
    E_{aprx} = \frac{\mathrm{Tr}(HA_{aprx}\rho A_{aprx}^\dagger)}{\mathrm{Tr}(A_{aprx}\rho A_{aprx}^\dagger)},
    \label{eq:approx_energy_using_A}
\end{equation}
is computed classically and serves as the reference for the estimators. It carries a truncation bias $E_{aprx} - E_{FCI}$ that is common to \textsc{HERMES} and PSA-LCU and therefore does not affect their comparison; its magnitude is at most $\approx 0.03$~Hartree for $\mathrm{H}_2$ and at most $0.02$~Hartree for $\mathrm{LiH}$ (Figs.~\ref{fig:h2_energy} and~\ref{fig:lih_energy}). Direct classical evaluation of $E_{aprx}$ becomes intractable as the system size grows, as it requires exact manipulation of an exponentially large state.

Two errors are reported for an estimate $\hat{E}$: $\varepsilon_{aprx} = |\hat{E} - E_{aprx}|$, which isolates the accuracy of the expectation-value estimator, and $\varepsilon_{FCI} = |\hat{E} - E_{FCI}|$, which additionally includes the truncation bias and shows the end-to-end deviation from the exact ground state. The estimates $\hat{E}$ are finite-sample estimators that carry statistical error from Monte Carlo sampling and shot noise (and, in noisy simulations, hardware noise) relative to $E_{aprx}$, and they are therefore not constrained to lie above $E_{FCI}$.

\paragraph{Simulation settings.}
All quantum circuits are constructed using PennyLane~\cite{bergholm2018PennylaneSoftware}. All simulations are performed on an HPC node with a 128-core AMD EPYC-Milan processor (one hardware thread per core) and $157$~GiB of system memory, of which approximately $150$~GiB was available to the computation. Both methods use identical sampling parameters: $10{,}000$ Monte Carlo samples and $10{,}000$ measurement shots per circuit. Ideal results correspond to noiseless, finite-shot simulations, whereas noisy results are obtained by simulating the circuits on the IBM \texttt{FakeSherbrooke} backend, which provides a noise model based on a calibration snapshot of the IBM \texttt{Sherbrooke} processor, with the circuits transpiled at optimization level~$3$~\cite{qiskit2024}.

\subsection{H$_2$ ground-state energy}
\label{subsec:h2}
 
Fig.~\ref{fig:h2_energy} shows the ground-state energy of $\mathrm{H}_2$ for five bond lengths from $0.50$ to $1.02$~\AA{}. The $E_{aprx}$ values (\ref{eq:approx_energy_using_A}) follow the expected potential-energy profile, with a minimum near the equilibrium bond length of $0.74$~\AA{}, and remain close to $E_{FCI}$.
 
In the ideal, noiseless, setting, both \textsc{HERMES} and PSA-LCU reproduce $E_{aprx}$ closely over the entire range (Fig.~\ref{fig:h2_error}). \textsc{HERMES} thus attains the accuracy of PSA-LCU without the ancilla qubit and the controlled gates that PSA-LCU requires.
 
Under noise, the two methods separate sharply. \textsc{HERMES} retains the shape of the potential-energy curve, including a minimum (shifted from $0.74$ to $0.90$~\AA{}), and its error is confined to the shortest bond lengths, decaying rapidly as the bond length increases (Figs.~\ref{fig:h2_energy} and~\ref{fig:h2_error}). Noisy PSA-LCU, in contrast, fails over the entire range: its error is large at every geometry, its estimates are positive at the shortest bond lengths, and no potential-energy minimum is recovered. The residual \textsc{HERMES} error at $0.50$~\AA{} ($\approx 0.45$~Hartree) shows, however, that \textsc{HERMES} is not immune to noise at the shortest bond length.
 
With respect to $E_{FCI}$ (Fig.~\ref{fig:h2_fci_error}), the noiseless errors of both methods are governed by the truncation bias and are largest at $0.50$~\AA{}, whereas the noisy errors follow the same trend as in Fig.~\ref{fig:h2_error}, so that the advantage of \textsc{HERMES} under noise persists with respect to the exact ground state.

\subsection{LiH ground-state energy}
\label{subsec:lih}
 
Fig.~\ref{fig:lih_energy} shows the ground-state energy of $\mathrm{LiH}$ for five bond lengths from $0.99$ to $2.01$~\AA{}. This is a substantially larger problem than $\mathrm{H}_2$: the PSA-LCU implementation requires a $13$-qubit circuit, including the ancilla. Both $E_{aprx}$ and $E_{FCI}$ are lowest at $1.57$~\AA{}, and $E_{aprx}$ lies slightly above $E_{FCI}$ at all geometries owing to the truncation bias.

In the ideal, noiseless setting, \textsc{HERMES} tracks $E_{aprx}$ closely throughout the scan (Fig.~\ref{fig:lih_error}). PSA-LCU follows the overall profile, but its error is larger than that of \textsc{HERMES} at four of the five bond lengths, the exception being $2.01$~\AA{}. The largest gap occurs at $0.99$~\AA{}, where the PSA-LCU error is $\approx 0.12$~Hartree against $\approx 0.01$~Hartree for \textsc{HERMES}. The PSA-LCU estimates also lie below $E_{aprx}$ at most geometries.

Under noise, \textsc{HERMES} continues to reproduce the energy profile (Figs.~\ref{fig:lih_energy} and~\ref{fig:lih_error}), with noise having a limited effect at most geometries. The exception is $0.99$~\AA{}, where the error increases to approximately $0.10$~Hartree, comparable to the noiseless PSA-LCU error and consistent with the sensitivity observed at the shortest $\mathrm{H}_2$ bond length. In contrast, the noisy PSA-LCU simulation was terminated after approximately $570$ circuit executions due to its substantially higher memory and computational-time requirements. Consequently, a complete noisy comparison with PSA-LCU could not be obtained within the available computational resources.

With respect to $E_{FCI}$ (Fig.~\ref{fig:lih_fci_error}), the noiseless \textsc{HERMES} deviations are of the order of the truncation bias. The noiseless PSA-LCU energies, in contrast, fall below $E_{FCI}$ at four of the five geometries, most strongly at $0.99$~\AA{}, which indicates estimator error. The \textsc{HERMES} estimates remain above $E_{FCI}$ except at $0.99$~\AA{}, where the noisy estimate (and marginally the noiseless one) lies below it.

\subsection{Discussion}
\label{subsec:discussion}
 
Table~\ref{tab:mae} lists the mean absolute error (MAE) of $\varepsilon_{aprx}$ and $\varepsilon_{FCI}$ over the five bond lengths of each molecule. Evaluating the methods over a range of bond lengths, rather than at a single equilibrium geometry, provides a more stringent assessment of expectation-value estimation. In the ideal, noiseless setting, the MAEs of \textsc{HERMES} and PSA-LCU for $\mathrm{H}_2$ agree to within $2\%$ with respect to both $E_{aprx}$ and $E_{FCI}$, whereas for $\mathrm{LiH}$ \textsc{HERMES} reduces the MAE by a factor of $4.6$ with respect to $E_{aprx}$ and by a factor of $2.3$ with respect to $E_{FCI}$. Under noise, \textsc{HERMES} reduces the MAE for $\mathrm{H}_2$ by factors of $7.9$ and $7.6$ with respect to $E_{aprx}$ and $E_{FCI}$, respectively, compared with PSA-LCU, while for $\mathrm{LiH}$ only \textsc{HERMES} yields results.

\begin{table}[t]
\centering
\caption{Mean absolute error (MAE) in Hartree over the five bond lengths of each molecule. A dash indicates that the noisy PSA-LCU simulation for LiH could not be completed with the available resources.}
\label{tab:mae}
\small
\setlength{\tabcolsep}{3pt}
\begin{tabular}{llcccc}
\hline
Molecule & Method &
\multicolumn{2}{c}{$\varepsilon_{\mathrm{aprx}}$} &
\multicolumn{2}{c}{$\varepsilon_{\mathrm{FCI}}$} \\
& & Ideal & Noisy & Ideal & Noisy \\
\hline
\multirow{2}{*}{H$_2$} & \textsc{HERMES} & 0.00111 & 0.124 & 0.00670 & 0.1310 \\
                       & PSA-LCU         & 0.00109 & 0.984 & 0.00675 & 0.9910 \\
\hline
\multirow{2}{*}{LiH}   & \textsc{HERMES} & 0.00756 & 0.027 & 0.01290 & 0.0308 \\
                       & PSA-LCU         & 0.03510  & --     & 0.03000 & --     \\
\hline
\end{tabular}
\end{table}

The improved noisy performance of \textsc{HERMES} can be understood from the gate content and depth of the executed circuits. Table~\ref{tab:circuit_resources} lists, for each method, the maximum number of single-qubit gates, the maximum number of two-qubit gates, and the maximum circuit depth over all circuits executed at the equilibrium bond length of each molecule, both for the ideal circuits and for the same circuits after transpilation to the \texttt{FakeSherbrooke} backend at optimization level~3 (native gate set \{\texttt{ECR}, \texttt{ID}, \texttt{Rz}, \texttt{SX}, \texttt{X}\}). The transpiled circuits are those executed in the noisy simulations. The \textsc{HERMES} circuits contain no two-qubit gates, both before and after transpilation, and after transpilation they have a depth of at most $3$ and contain at most $12$ and $36$ single-qubit gates for $\mathrm{H}_2$ and $\mathrm{LiH}$, respectively. In contrast, the PSA-LCU circuits for $\mathrm{H}_2$ contain up to $8$ two-qubit gates in the ideal case, which increases to $14$ after transpilation, with the maximum depth increasing from $11$ to $66$. The overhead is considerably larger for $\mathrm{LiH}$, where the transpiled PSA-LCU circuits contain up to $96$ two-qubit gates and have a depth of up to $279$. This growth arises from the translation of the controlled Pauli operations into the native gate set and from the additional gates required to route interactions between the ancilla and the system qubits under the limited connectivity of the device. Since two-qubit gates have substantially higher error rates than single-qubit gates, and since decoherence accumulates with circuit depth, the absence of two-qubit gates and the very small depth of the \textsc{HERMES} circuits provide a natural explanation for its much smaller noise-induced error, whereas the large two-qubit gate count and depth of the PSA-LCU circuits are consistent with its large noisy error (Table~\ref{tab:mae}, Figs.~\ref{fig:h2_results} and~\ref{fig:lih_results}). The same overhead also makes the noisy simulation of PSA-LCU considerably more demanding for $\mathrm{LiH}$.

\begin{table}[b]
\centering
\caption{Maximum numbers of single-qubit (1Q) gates, two-qubit (2Q) gates, and circuit depth among all executed circuits at the equilibrium bond length ($0.74$~\AA\ for $\mathrm{H}_2$, $1.57$~\AA\ for $\mathrm{LiH}$), for the ideal circuits and after transpilation to \texttt{FakeSherbrooke}.}
\label{tab:circuit_resources}
\small
\setlength{\tabcolsep}{3pt}
\begin{tabular}{llcccc}
\hline
Molecule & Quantity &
\multicolumn{2}{c}{\textsc{HERMES}} &
\multicolumn{2}{c}{PSA-LCU} \\
& & Ideal & Transpiled & Ideal & Transpiled \\
\hline
\multirow{3}{*}{H$_2$} & 1Q gates & 6  & 12 & 7  & 89  \\
                       & 2Q gates & 0  & 0  & 8  & 14  \\
                       & Depth    & 2  & 3  & 11 & 66  \\
\hline
\multirow{3}{*}{LiH}   & 1Q gates & 16 & 36 & 7  & 469 \\
                       & 2Q gates & 0  & 0  & 21 & 96  \\
                       & Depth    & 2  & 3  & 24 & 279 \\
\hline
\end{tabular}
\end{table}

Beyond accuracy and circuit complexity, the two methods differ in how their measurement workload scales with the number of Monte Carlo samples. In PSA-LCU, the measurement cost increases with the number of samples because each sample requires a separate set of measurement circuits. In contrast, \textsc{HERMES} uses Pauli deduplication and commuting-group partitioning to avoid repeated quantum measurements, shifting the associated computational cost to classical pre- and post-processing. Although this introduces a classical overhead that grows with the number of Pauli terms in $A$, the trade-off is favorable when quantum execution is the dominant computational bottleneck. Under identical computational resources, \textsc{HERMES} remained tractable for the noisy $\mathrm{LiH}$ simulation, whereas PSA-LCU could not complete the corresponding simulation within the same computational budget. Overall, \textsc{HERMES} provides comparable or improved accuracy with simpler circuits and is more robust and tractable under noise than PSA-LCU.


\section{Conclusions}
\label{sec:conclusion}

We have presented \textsc{HERMES}, a hybrid classical--quantum algorithm for estimating $\mu = \mathrm{Tr}(OA\rho A^\dagger)/\mathrm{Tr}(A\rho A^\dagger)$ that shifts the entire quantum processing of $A$ to the classical domain. By expanding $A$ and $O$ in the Pauli basis and sampling from the coefficient distribution of $A$, \textsc{HERMES} rewrites $\mu$ as a ratio of weighted sums of Pauli expectation values on the unmodified input state $\rho$. The resulting quantum circuit consists solely of state preparation and Clifford basis rotations, with no ancilla qubits, no controlled unitaries (with QWC grouping) and no quantum implementation of $A$. A two level measurement reduction scheme bounds the number of distinct measurement settings independently of the Monte Carlo sample count, in contrast to existing randomized LCU methods such as single-ancilla LCU, whose circuit executions grow linearly with the sample count. Simulation results showed that, for ground-state estimation of $\mathrm{H}_2$ and $\mathrm{LiH}$, \textsc{HERMES} (with QWC grouping) achieves comparable or improved accuracy relative to PSA-LCU baseline with circuits that contain no two-qubit gates and have a much smaller depth.  Furthermore, \textsc{HERMES} exhibits greater robustness to hardware noise and remains computationally feasible for $\mathrm{LiH}$. These properties make \textsc{HERMES} attractive for NISQ and early fault-tolerant devices, where ancilla qubits and multi-qubit gates are costly, although this comes with a classical overhead that grows with the number of Pauli terms in $A$.

In addition to ground-state estimation, \textsc{HERMES} is applicable to other LCU-based expectation value problems, such as Hamiltonian simulation~\cite{berry2015simulating_truncated_Taylor_series} and quantum linear systems~\cite{harrowHHL2009quantum}, provided that $O$ and $A$, or a polynomial approximation of $A$, admit Pauli expansions with a manageable number of terms; the same sampling and measurement pipeline then applies. A key open research direction is the derivation of a rigorous convergence guarantee for the \textsc{HERMES} estimator, providing explicit sample-complexity bounds. Exploring general commuting groups could further reduce the number of measurement settings, at the cost of additional two-qubit gates in the basis-rotation circuits. Moreover, since Pauli partitioning is part of the classical preprocessing, advances in Pauli-partitioning algorithms, such as the method proposed in~\cite{reggio2024fast_partition_Paulis}, would directly benefit \textsc{HERMES}.

\section*{ACKNOWLEDGEMENTS}
The authors acknowledge Dr. Nagendra Nagaraja, Lakshya Priyadarshi and the management of QpiAI India Private Limited for their support in providing the computational resources required for this work.


\section*{Data Availability}
The data that support the findings of this article are available upon reasonable request from the corresponding author. The software used to generate the numerical results is proprietary and is not publicly available.

\section*{Author Contributions}
HHSC conceptualized and developed the algorithms. TM and HHSC developed the software. TM conducted numerical simulations and analysis. HHSC and TM prepared the manuscript. All authors reviewed and approved the final version of the manuscript.

\section*{Competing Interests}

HHSC is an inventor in a patent application, related to the methods described in this work, filed by QpiAI India Private Limited.

\section*{AI Disclosure}
All ideas and writing are due to the authors of this work. GPT-6 and Claude Sonnet 5.5 were used to proofread the manuscript and suggest minor edits.


\bibliography{Arxiv}

\end{document}